\documentclass{article}

\usepackage{PRIMEarxiv}

\usepackage[utf8]{inputenc} 
\usepackage[T1]{fontenc}    
\usepackage{hyperref}       
\usepackage{url}            
\usepackage{booktabs}       
\usepackage{amsfonts}       
\usepackage{nicefrac}       
\usepackage{microtype}      
\usepackage{lipsum}
\usepackage{fancyhdr}       
\usepackage{graphicx}       
\usepackage{amsmath}
\usepackage{subfig}
\usepackage{makecell}

\graphicspath{{media/}}     

\title{Retracing the Process of Translation: Proteome-wide mapping of stable transcriptomic predictors of protein abundance in cancer cell lines
}

\author{
  Johannes Schlüter \\
  Universität Bielefeld \\
  Bielefeld\\
  \texttt{j.schlueter@uni-bielefeld.de} \\
   \And
   Alexander Schönhuth \\
  Universität Bielefeld  \\
  Bielefeld\\
  \texttt{aschoen@cebitec.uni-bielefeld.de} \\
}

\begin{document}
\maketitle

\begin{abstract}
\subsubsection*{Background:}  
Understanding the relationship between gene expression and protein abundance is central to molecular and systems biology. While gene expression reflects transcriptional activity, proteins are the functional molecules that determine cellular phenotypes. However, numerous post-transcriptional and translational regulatory layers complicate this relationship, and prior studies have reported only weak to moderate correlations between RNA and protein levels. Predicting protein abundance from transcriptomic data remains challenging, but it is a valuable goal for biological insight, especially when proteomic data is limited or unavailable.

\subsubsection*{Results:}  
In this study, we applied a large-scale, Ridge regression-based feature selection strategy to identify predictive gene expression features for each of 8,423 proteins across 940 cancer cell lines. To our knowledge, this is the first work to perform such comprehensive protein-wise feature selection at this scale. Our analysis revealed both globally predictive and context-specific gene features. These included biologically meaningful modules such as immune-related genes, HOX transcription factor targets, and cytoskeletal components. The models identified stable candidate gene–protein associations that remained interpretable at the level of individual proteins and recurrent transcriptomic predictor patterns.

\subsubsection*{Conclusion:}  
Our approach enables interpretable modeling of protein expression from transcriptomic data and provides insight into transcriptomic features associated with protein abundance. This framework may support hypothesis generation, protein imputation in incomplete datasets, and deeper understanding of post-transcriptional regulation in cancer biology.
\end{abstract}


\centering  
\includegraphics[width=\textwidth,height=0.75\textheight,keepaspectratio]{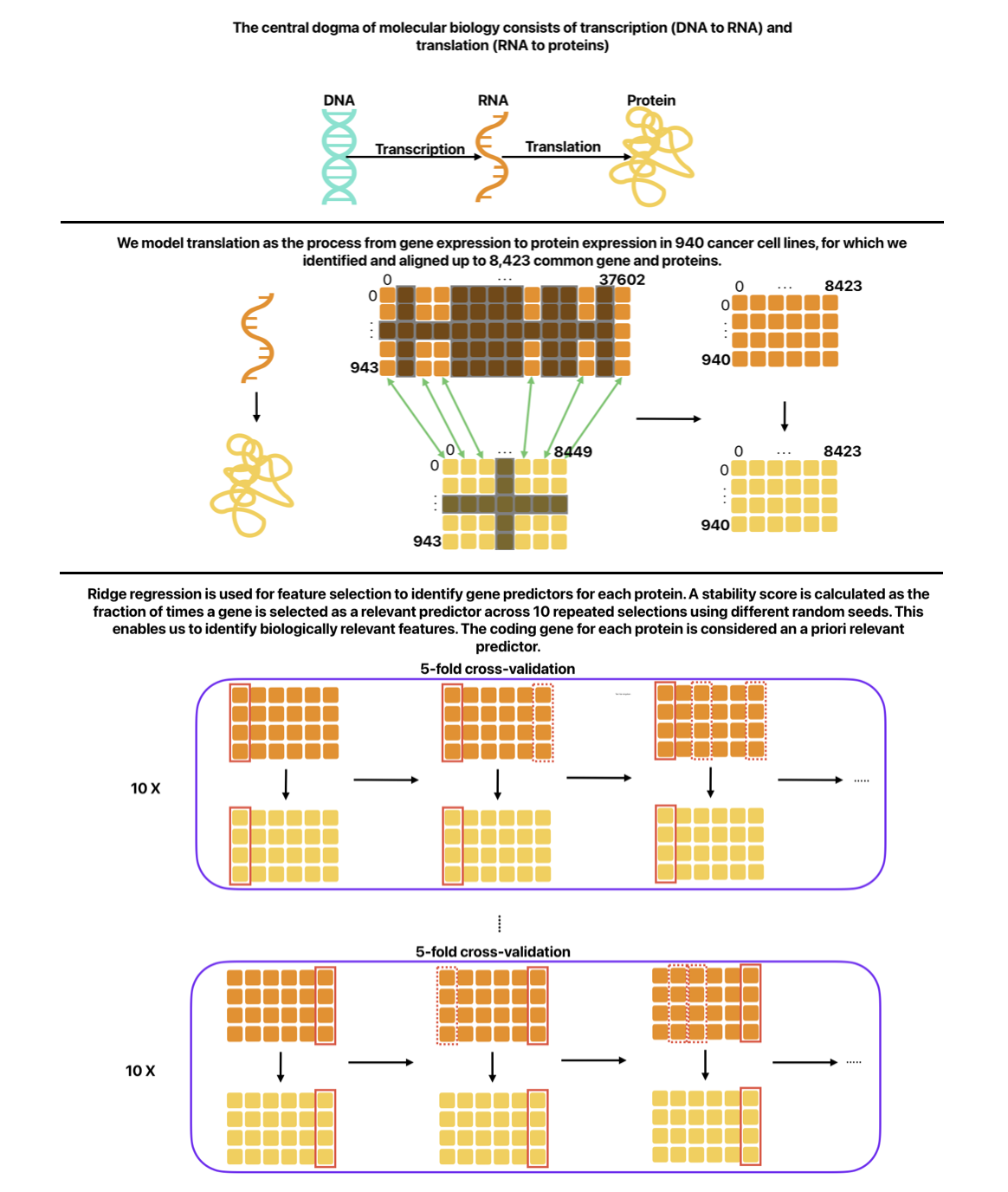}
\label{fig_graphical_abstract}



\section{Introduction}

\begin{figure}[thb]
    \centering
    \includegraphics[width=\linewidth]{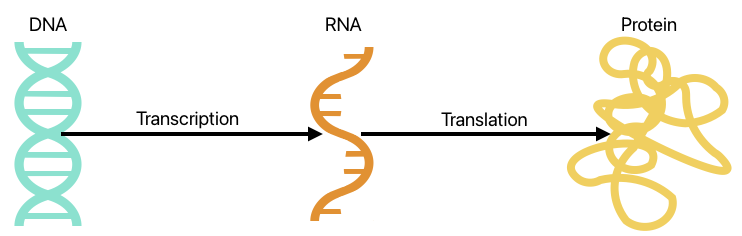}
    \caption{Central dogma of molecular biology. In theory, RNA is translated into proteins, thus, determining protein abundance.}
    \label{fig:central_dogma}
\end{figure}

The central dogma of molecular biology describes the directional flow of genetic information from DNA to RNA to proteins (Figure~\ref{fig:central_dogma})~\cite{Koonin2012, ROCHA201859}. While this process suggests a linear relationship, it is now well established that gene expression and protein abundance are influenced by multiple layers of regulation (e.g., transcriptional control, post-transcriptional and translational mechanisms, protein stability)~\cite{Vogel2012, Srivastava2022, DOUGCHUNG2013369}. Consequently, mRNA levels do not necessarily predict protein levels with high accuracy~\cite{Nicolet2022, REVIEWgenes14112065}.

Gene expression is typically measured using transcriptomic technologies such as RNA sequencing (RNA-seq)~\cite{Wang2009}, while protein abundances are quantified by proteomic methods including Reverse Phase Protein Arrays (RPPA)~\cite{Byron2020, RPPACoarfa2021-ue} and mass spectrometry (e.g., DIA-MS, DDA-MS)~\cite{Aebersold2016, DIADoerr2015}. Proteins undergo extensive post-transcriptional and post-translational modifications that affect their final abundance, making the inference of protein levels from RNA data inherently challenging~\cite{Vogel2012, Kozoriz2025}.

Previous studies investigating the correlation between gene and protein expression reported only moderate agreement (Pearson’s $r$ typically ranging between 0.3 and 0.5)~\cite{Nicolet2022, Srivastava2022, DOUGCHUNG2013369}. This limited predictability indicates that transcriptomic information alone is insufficient to fully reconstruct the proteome~\cite{Vogel2012, REVIEWgenes14112065}.

Traditional approaches for predicting protein abundance from gene expression data have employed statistical or classical machine learning models such as linear regression, random forests, and shallow neural networks~\cite{Li2019}. Ferreira et al., for instance, used ridge regression and random forests to predict protein abundances from codon usage and achieved $R^2$ values between 0.62 and 0.73~\cite{Li2019}.

More recently, deep learning methods such as convolutional and recurrent neural networks have improved prediction performance by capturing complex, nonlinear patterns in biological data~\cite{Schwehn2025, Zhou2020}. Barzine et al. demonstrated this by using a deep neural network to predict protein levels, achieving $R^2$ scores up to 0.631~\cite{Barzine2020-fo}. Their model was trained on a subset of proteins with known transcriptomic and proteomic measurements and used to impute missing protein abundances within the same cell line, based on gene expression and functional annotations (e.g., Gene Ontology terms). While this is valuable for reconstructing incomplete proteomic profiles, it differs fundamentally from our approach: Barzine et al. focused on within-cell-line imputation, whereas our method identifies transcriptomic predictors across cell lines and enables interpretable gene–protein associations at the population level.

\begin{figure}[thb]
    \centering
    \includegraphics[width=\linewidth]{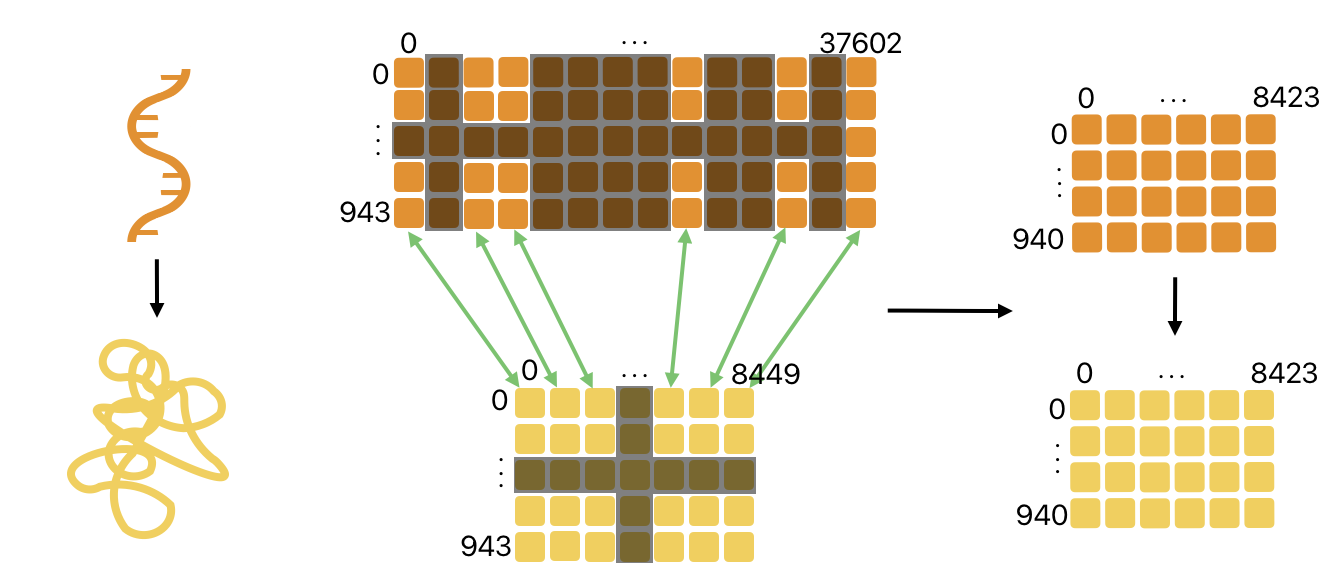}
    \caption{Translation represented as gene expression values being translated to protein expression for 940 cancer cell lines. Here, the alignment of gene expression values to protein abundances values is shown. The original data set contains 37,602 gene expression values and 8,449 proteins after cleaning empty or broken entries. We computed the common genes/proteins between both matrices and removed empty cell lines. This yields 940 cell lines with 8,423 gene-protein-pairs.}
    \label{fig:translation}
\end{figure}

Our study represents a fundamentally different goal and starting point compared to prior work: Rather than predicting entire proteomes at once or imputing missing protein abundances, we ask which genes contribute most to the expression of individual proteins. To achieve this, we apply greedy feature selection using ridge regression to identify stable mRNA predictors of protein abundance~\cite{Hanhart2024}. This enables us to explore not only predictive accuracy but also the underlying biological mechanisms that connect gene expression to protein levels in cancer cell lines (Figure~\ref{fig:translation}).


\begin{figure}[thb]
    \centering
    \includegraphics[width=\linewidth]{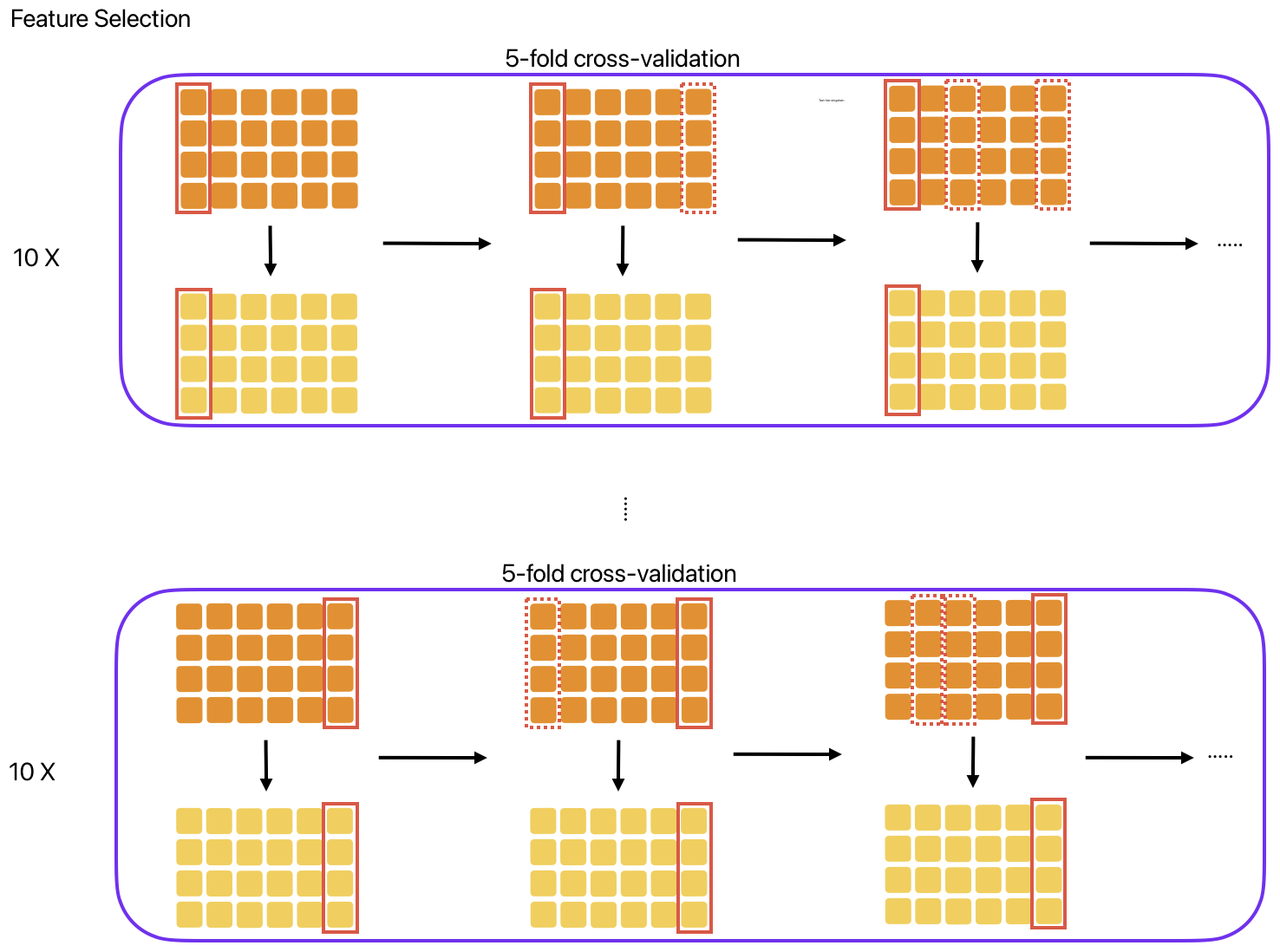}
    \caption{Forward feature selection for each protein separately with 10 random seeds starting with the \textit{cis-feature} of the corresponding protein. Our Ridge regression is performed for each protein starting with its cis-feature. Then, the next feature is selected based on its contribution to the $R^2$ together with the cis-feature. This is done until one of two stopping criteria are triggered: i) a maximum of $n-1 = 30$ features and the \textit{cis-feature} ($n = 31$ in total) was reached or ii) the incremental gain in $R^2$ dropped below a threshold of 0.0001. This feature selection is done for 10 different random seed for 5-fold cross-validation for each protein separately.}
    \label{fig:feature_selection}
\end{figure}

For each protein, we start with its \textit{cis-feature}. The \textit{cis-feature} is the gene expression of the gene that encodes for the corresponding protein. This leverages biological prior knowledge by prioritizing the transcript corresponding to the target protein. To ensure importance and interpretability of selected features, we perform our feature selection with 10 different random seeds (Figure~\ref{fig:feature_selection}). Setting a random seed ensures that the random shuffling of data before cross-validation is consistent across runs, making the results reproducible. By using 10 different random seeds, we ensure that data splits differ across runs.
This enables us to discover stable genetic features per protein and globally recurrent transcriptomic predictors. Through stability analysis and enrichment analysis of resulting feature sets we underline our biologically motivated approach yielding stability and reproducibility of candidate transcriptomic signatures associated with protein abundance in cancer.
We employ linear Ridge regression to ensure high interpretability while avoiding the black-box nature of more complex models. Its $L_2$ regularization makes it particularly suited for high-dimensional, collinear gene expression data, where it stabilizes coefficient estimates without discarding correlated features~\cite{sokolov2016}. Unlike sparse models such as Lasso, which may arbitrarily exclude informative genes, Ridge provides a more stable and reproducible foundation for identifying gene–protein associations. This regularization also mitigates overfitting, helping the model avoid capturing noise inherent in biological data.
Furthermore, we demonstrate strong explainability for several hundred proteins through gene expression data only while minimizing noise introduced by irrelevant or corrupted gene expression features.

In summary, our work aims to systematically uncover stable, interpretable gene–protein associations across the proteome using a scalable, biologically motivated framework. By analyzing thousands of proteins in hundreds of cancer cell lines and combining ridge regression with feature selection stability, we provide new insights into the extent to which transcriptomic data can explain protein abundance. This study not only advances the systematic characterization of transcriptome–proteome coupling in cancer, but also lays the foundation for robust biomarker discovery and multi-omics integration in future research.

\section{Results}

\subsection{Stability of selected gene features across proteins}

To evaluate the robustness of selected features, we performed greedy feature selection with 10 different random seeds per protein as described in details in our \textit{Methods} section. For each gene feature, we computed a \textit{stability score}, defined as the fraction of runs in which the feature was selected.

\begin{figure}[ht]
    \centering
    \includegraphics[width=\linewidth]{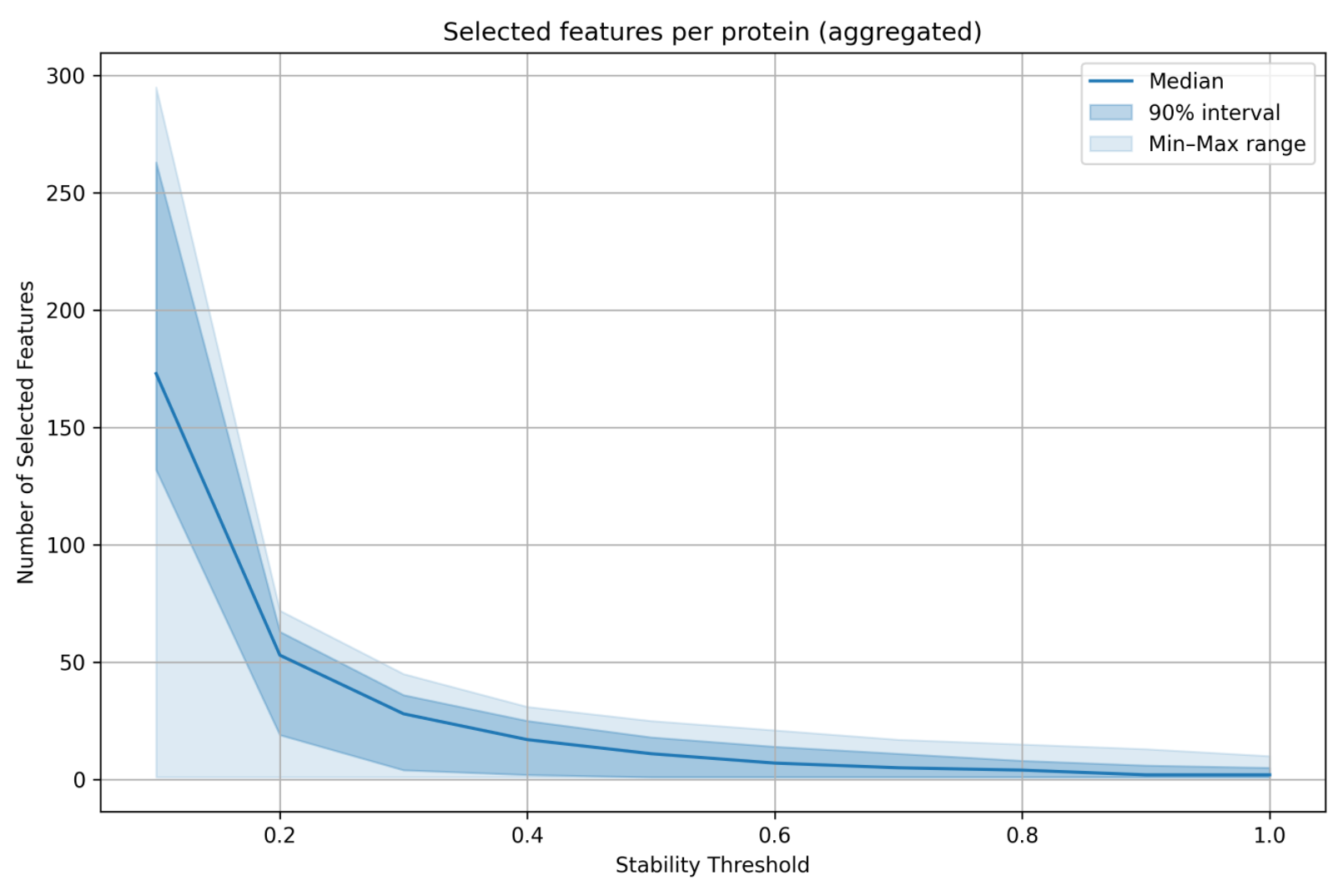}
    \caption{Number of selected features per protein across \textit{stability scores}. The number of features selected decreases from lower to higher \textit{stability scores}.}
    \label{fig:stability_summary}
\end{figure}

Figure~\ref{fig:stability_summary} summarizes the number of stable features per protein at various \hyperref[sec:StabilityScore]{\textit{stability scores}}.
For example, a feature with a \textit{stability score} of 1.0 was selected in all 10 performed runs, indicating strong robustness and high relevance for the corresponding protein. In contrast, a feature selected in only one out of 10 runs yields a \textit{stability score} of 0.1, suggesting lower robustness and a higher likelihood of being introduced by noise.
As the \textit{stability score} increased from 0.1 to 1.0, the number of retained features decreased markedly for all proteins. The maximal number of selected features for a protein was 295 while the maximal theoretical number for 10 random seeds and stopping criteria i) is 300. 

At the low threshold of 0.1, between 132 and 263 features were selected for most proteins (Table~\ref{tab:stability_feature_distribution}). However, raising the threshold to 0.2 already reduces the number of stable features dramatically—indicating that only about one third of the features selected at 0.1 were consistently selected in at least two out of ten runs. This trend of decreasing feature count continues with increasing stability, though the rate of decline flattens for thresholds above 0.5.

These observations highlight a trade-off between the number of explanatory features and their robustness. While lower thresholds capture a broader set of potentially relevant features, only a smaller subset demonstrates consistency across different random seeds and training splits. Selecting a meaningful \textit{stability score}, thus, involves balancing comprehensiveness with reproducibility.

\begin{table}[ht]
\centering
\caption{Distribution of selected features per protein across different \textit{stability scores}. }
\label{tab:stability_feature_distribution}
\begin{tabular}{lrrrrr}
\toprule
\textbf{Threshold} &

\makecell{\textbf{Min}} &
\makecell{\textbf{Lower}\\\textbf{Quartile}} &
\makecell{\textbf{Median}} &
\makecell{\textbf{Upper}\\\textbf{Quartile}} &
\makecell{\textbf{Max}} \\
\midrule
0.1 & 1 & 132.0 & 173.0 & 263.0 & 295.0 \\
0.2 & 1 & 19.0 & 53.0  & 63.0  & 72.0 \\
0.3 & 1 & 4.0  & 28.0  & 36.0  & 45.0 \\
0.4 & 1 & 2.0  & 17.0  & 25.0  & 31.0 \\
0.5 & 1 & 1.0  & 11.0  & 18.0  & 25.0 \\
0.6 & 1 & 1.0  & 7.0   & 14.0  & 21.0 \\
0.7 & 1 & 1.0  & 5.0   & 11.0  & 17.0 \\
0.8 & 1 & 1.0  & 4.0   & 8.0   & 15.0 \\
0.9 & 1 & 1.0  & 2.0   & 6.0   & 13.0 \\
1.0 & 1 & 1.0  & 2.0   & 5.0   & 10.0 \\
\bottomrule
\end{tabular}
\end{table}

\begin{figure}[ht]
    \centering
    \includegraphics[width=\linewidth]{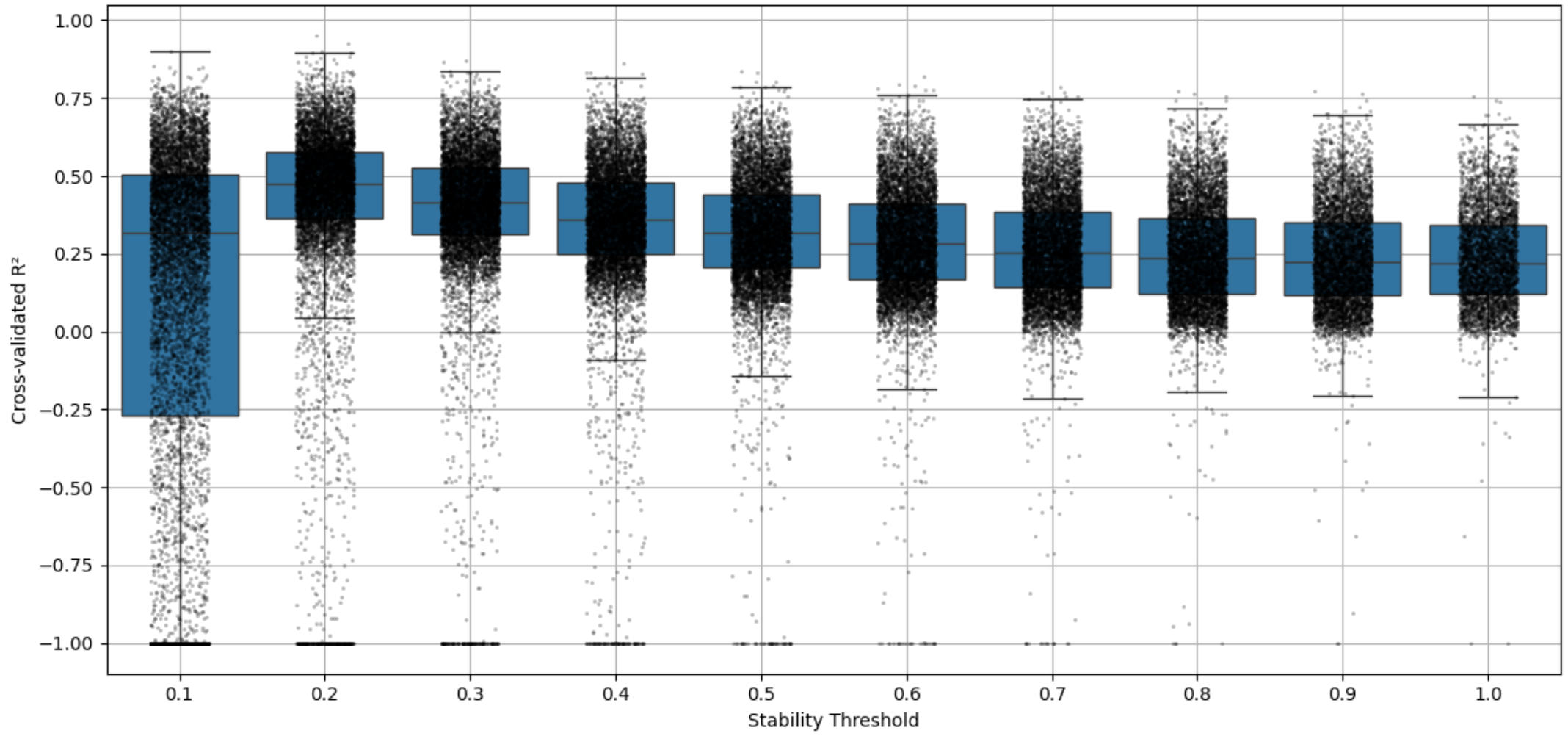}
    \caption{Distribution of cross-validated $R^2$ scores for 8,423 proteins across different \textit{stability score} thresholds (0.1 to 1.0). Each point represents the performance of a ridge regression model trained with features selected at the given threshold. Lower thresholds include more noisy features, resulting in greater variance and more poor-performing models. Higher thresholds reduce variance and outliers, favoring more stable and interpretable feature sets. The plot illustrates the trade-off between predictive power and feature robustness. Notably, \textit{stability score} $0.1$ achieves more negative $R^2$ scores and fewer very high scores.}
    \label{fig:R2DistributionStabilityThreshold}
\end{figure}

\subsection{Transcriptome-based predictability of protein abundance}

\begin{figure}
    \centering
    \includegraphics[width=0.67\linewidth]{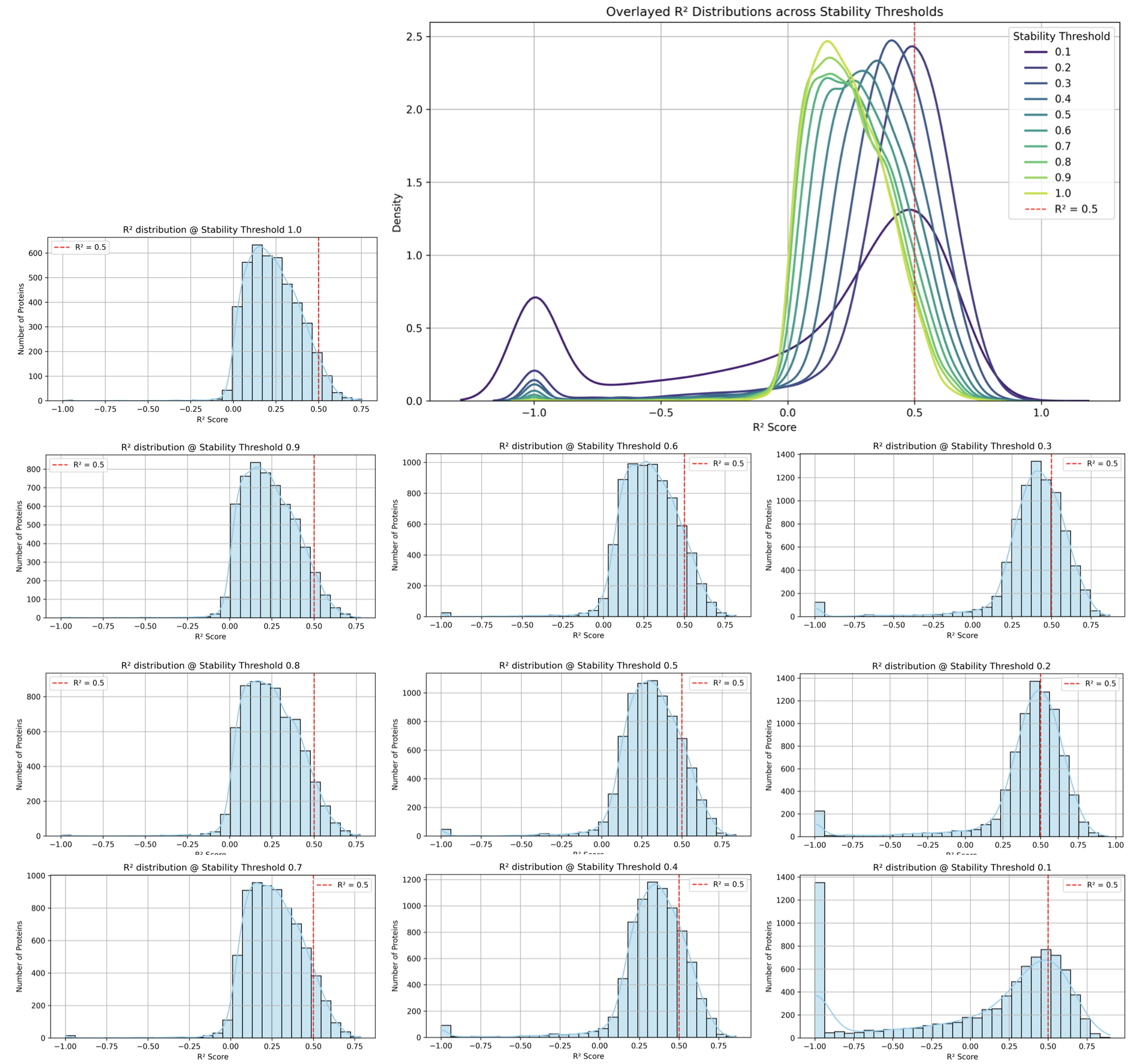}
    \caption{Distribution of cross-validated $R^2$ scores for all proteins per \textit{stability score}. Notably, a \textit{stability score} of 0.1 is associated with more negative $R^2$ scores and fewer high $R^2$ scores ($\geq 0.5$) than \textit{stability score} 0.2. However, very high \textit{stability scores} ($\geq 0.8$) show a very small number of negative $R^2$ scores, but mostly achieve $R^2$ scores between $0$ and $0.5$. 
    \textit{stability score} $0.6$ as a trade-off between high explanatory capacity (\textit{stability score}$=0.2$) and biological robustness (\textit{stability score}$=1.0$) shows a minimal number of negative $R^2$ scores while introducing very high explanatory capacities with by majority consensus robustly selected features.}
    \label{fig:r2_hist}
\end{figure}

After applying cross-validated ridge regression models for each of the 8{,}423 proteins to select interpretable biological features, we applied ridge regression again using only prior selected features. Based on 5-fold cross-validation, we observed a wide range of predictability across proteins. Figure~\ref{fig:r2_hist} shows the distribution of $R^2$ scores across all proteins.

Table~\ref{tab:stability_r2_distribution} summarizes the $R^2$ per \textit{stability score}. Here, it is important to notice that we did not find stable features for each of the 8,423 proteins for all \textit{stability scores}. For example, features with a \textit{stability score} of $1.0$ were found for only about half of all proteins (4,339). For 257 proteins, we were not able to report any selected features. This was due to insufficient data. $R^2 \leq -1$ were clipped to $-1$ to avoid too great impacts of negative values and to provide good readability as $R^2$ scores are not limited in the negative direction.
However, a substantial subset of proteins could be well explained from transcriptomic data alone:  
Looking at \textit{stability score} 0.2, we notice that 3,555 proteins are explained with $R^2$ scores of 0.5 or greater. The number of well explained proteins decreases with greater \textit{stability scores} while for \textit{stability score} 0.1  the number decreases only slightly. Moreover, for very well explainable proteins ($R^2 >= 0.8$), features selected with \textit{stability score} 0.1 (15) perform almost as well as features selected with \textit{stability score} 0.3 (17). Still, it is crucial to recognize that for \textit{stability score} 0.1 the number of proteins with very low explainability ($R^2<0$) is by far the largest (2,611). Additionally, we observe a decrease of very low performing proteins with higher \textit{stability scores}.
These values indicate that for a meaningful fraction of the proteome, gene expression data contains sufficient information to reconstruct protein abundance with moderate to high fidelity. While features with a \textit{stability score} of 0.2 seem to enable highest explainability for many proteins, it also introduces issues for a significant number of proteins (549). Thus, a higher \textit{stability score} introduces stability in terms of a small amount of not explainable proteins (83). This indicates an introduction of noisier feature and overfitting with lower \textit{stability scores}.
These phenomena are visualized in Figure~\ref{fig:r2_hist}.
However, proteins for which we did not achieve an $R^2$ of 0.5 or higher may require more complex, non-linear models or more comprehensive and reliable data to be accurately predicted. Nevertheless, the selected features—especially those with high \textit{stability scores}—are likely to be biologically relevant to a protein’s abundance, even if they do not fully explain it on their own.

\begin{table*}[ht]
\centering
\begin{tabular}{l|rrrrrrrrrr}
\toprule
\textbf{Stability Scores} & \textbf{0.1} & \textbf{0.2} & \textbf{0.3} & \textbf{0.4} & \textbf{0.5} & \textbf{0.6} & \textbf{0.7} & \textbf{0.8} & \textbf{0.9} & \textbf{1.0} \\
\midrule

\textbf{R\textsuperscript{2}}\\
\texttt{$< 0$}   & 2611 & 549  & 348  & 323  & 218  & 184  & 160  & 170  & 129  & 83  \\
0.0          & 278  & 85   & 103  & 149  & 342  & 636  & 990  & 1151 & 1111 & 812 \\
0.1          & 422  & 184  & 293  & 762  & 1292 & 1597 & 1595 & 1511 & 1345 & 1058 \\
0.2          & 632  & 494  & 1090 & 1645 & 1731 & 1640 & 1538 & 1444 & 1262 & 962 \\
0.3          & 931  & 1273 & 1879 & 1882 & 1708 & 1486 & 1325 & 1149 & 990  & 746 \\
0.4          & 1163 & 2026 & 1985 & 1628 & 1317 & 1139 & 973  & 818  & 643  & 456 \\
0.5          & 1149 & 1907 & 1525 & 1099 & 832  & 640  & 492  & 365  & 264  & 186 \\
0.6          & 741  & 1188 & 731  & 461  & 304  & 215  & 141  & 101  & 68   & 33  \\
0.7          & 224  & 405  & 188  & 94   & 54   & 26   & 17   & 8    & 8    & 3   \\
0.8          & 15   & 53   & 17   & 7    & 4    & 1    & 0    & 0    & 0    & 0   \\
0.9          & 0    & 2    & 0    & 0    & 0    & 0    & 0    & 0    & 0    & 0   \\
1.0          & 0    & 0    & 0    & 0    & 0    & 0    & 0    & 0    & 0    & 0   \\
\midrule
\textbf{SUM} & 8166 & 8166 & 8159 & 8050 & 7802 & 7564 & 7231 & 6717 & 5820 & 4339 \\
\bottomrule
\end{tabular}
\caption{Number of proteins per R\textsuperscript{2} and \textit{stability score}. "$< 0$" indicating any negative R\textsuperscript{2} score. The sum of evaluated proteins shrinks with greater \textit{stability scores}. For example, we found for 4,339 proteins features with a \textit{stability score} of 1.0.}
\label{tab:stability_r2_distribution}
\end{table*}

\subsection{Benchmarking Against Full Transcriptome Features}
\label{subsecFullTran}
To evaluate the influence of the selected input space, we benchmarked our approach using the full transcriptome (37,602 genes) instead of the restricted set of 8,423 focus-genes. We repeated the modeling procedure for 100 randomly chosen proteins, using the same ridge regression and cross-validation scheme as before.

The results confirm the validity of our design choice: in 90 out of 100 cases, the models based on focus-genes achieved higher $R^2$ scores than those based on the full transcriptome. When averaging across all \textit{stability scores}, focus-based models outperformed full-transcriptome models in 74 cases, compared to only 21 cases in the reverse direction. These findings indicate that our biologically grounded feature reduction improves both computational efficiency and predictive robustness, without sacrificing performance.

\subsection{Benchmarking Against Random Forests}
\label{subsecRandomForest}

To assess the performance of our interpretable linear model, we compared ridge regression to Random Forests, a widely used non-linear machine learning method. We conducted this benchmark on a randomly selected subset of 100 proteins using the same gene expression features and cross-validation protocol.

Despite the flexibility of Random Forests, ridge regression clearly outperformed them in this task. Across all 100 proteins, ridge regression achieved the highest cross-validated $R^2$. When averaged across stability score thresholds, ridge regression models yielded better performance in 88 cases, while Random Forests performed better in only 12 cases. These results underscore the suitability of ridge regression for high-dimensional, collinear gene expression data, providing both competitive predictive power and model interpretability.

\subsection{External Validation Using CCLE}

To assess the generalizability of our gene–protein associations, we performed an external validation using the independent CCLE proteomics dataset~\cite{CCLEBarretina2012}. 

Applying our GDSC2-derived predictor sets, we retrained ridge regression models on CCLE data and evaluated performance via Pearson correlation. At a stability score of 0.5, 3,679 of 7,155 protein models (51.4\%) achieved $PCC \geq 0.5$, while 2,403 (33.6\%) achieved $PCC \geq 0.6$. At a stability score of 0.7, 3,704 protein models (51.8\%) achieved $PCC \geq 0.5$, indicating that the predictive value of the transferred feature sets was largely retained under more stringent stability selection (Table~\ref{tab:stability_pcc_distribution}).
These results confirm that a substantial subset of our associations are reproducible across datasets, supporting their potential utility in translational settings.

\begin{table*}[ht]
\centering
\begin{tabular}{l|rrrrrrrrrr}
\toprule
\textbf{PCC} &
\textbf{0.1} & \textbf{0.2} & \textbf{0.3} & \textbf{0.4} &
\textbf{0.5} & \textbf{0.6} & \textbf{0.7} & \textbf{0.8} &
\textbf{0.9} & \textbf{1.0} \\
\midrule
$<0$          & 169  & 123  & 118  & 119  & 113  & 120  & 132  & 151  & 151  & 156 \\
$0.0$--$<0.1$ & 330  & 169  & 153  & 153  & 174  & 178  & 169  & 180  & 207  & 192 \\
$0.1$--$<0.2$ & 908  & 417  & 382  & 408  & 410  & 433  & 450  & 428  & 437  & 453 \\
$0.2$--$<0.3$ & 1528 & 743  & 691  & 673  & 674  & 660  & 671  & 674  & 660  & 656 \\
$0.3$--$<0.4$ & 1572 & 1139 & 1014 & 913  & 901  & 875  & 845  & 842  & 821  & 830 \\
$0.4$--$<0.5$ & 1258 & 1380 & 1242 & 1244 & 1204 & 1205 & 1184 & 1172 & 1171 & 1163 \\
$0.5$--$<0.6$ & 835  & 1266 & 1305 & 1288 & 1276 & 1271 & 1265 & 1269 & 1268 & 1263 \\
$0.6$--$<0.7$ & 449  & 1052 & 1100 & 1108 & 1100 & 1084 & 1097 & 1086 & 1090 & 1078 \\
$0.7$--$<0.8$ & 101  & 728  & 908  & 962  & 1003 & 1002 & 997  & 1001 & 994  & 1002 \\
$0.8$--$<0.9$ & 5    & 138  & 240  & 283  & 296  & 322  & 340  & 346  & 351  & 357 \\
$0.9$--$1.0$  & 0    & 0    & 2    & 4    & 4    & 5    & 5    & 6    & 5    & 5 \\
\midrule
\textbf{SUM} &
7155 & 7155 & 7155 & 7155 & 7155 &
7155 & 7155 & 7155 & 7155 & 7155 \\
\bottomrule
\end{tabular}

\caption{
Distribution of cross-validated Pearson correlation coefficients (PCCs)
for 7,155 proteins across different \textit{stability score} thresholds. GDSC-derived predictor sets were transferred to CCLE and evaluated using Ridge regression with 5-fold cross-validation. PCC values are grouped into non-overlapping intervals based on the original, unrounded correlation coefficients. All stability thresholds were evaluated on the same set of 7,155 proteins.
}
\label{tab:stability_pcc_distribution}
\end{table*}

\subsection{Gene–protein maps as interpretable signatures}

\begin{figure}[ht]

    \centering
    \includegraphics[width=\linewidth]{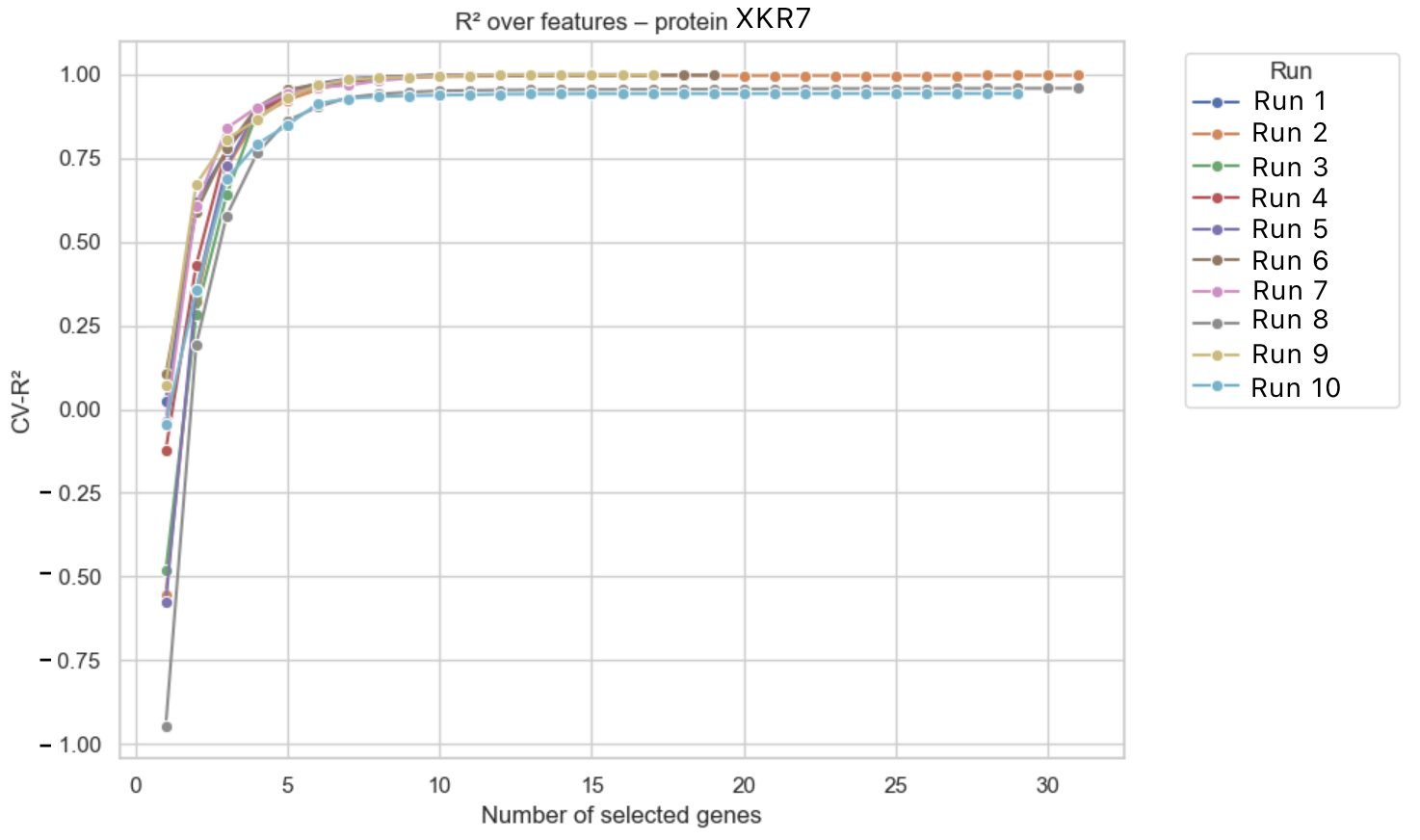}
    \caption{$R^2$ improvement with number of selected features across all 10 runs for well-predictable proteins \textbf{XKR7}. The improvement of the $R^2$ approximates a line after selecting 8-10 features for each run. Without selecting any features in addition to the cis-features, the model achieves $R^2$ scores below or around $0$. This indicates low explanatory relevance of the cis-feature, the one feature which needs to be involved by biological reasoning. Thus, great improvement is achieved by selecting only few relevant features.}
    \label{fig:xkr7}

\end{figure}

\begin{figure}[ht]

    \centering
    \includegraphics[width=\linewidth]{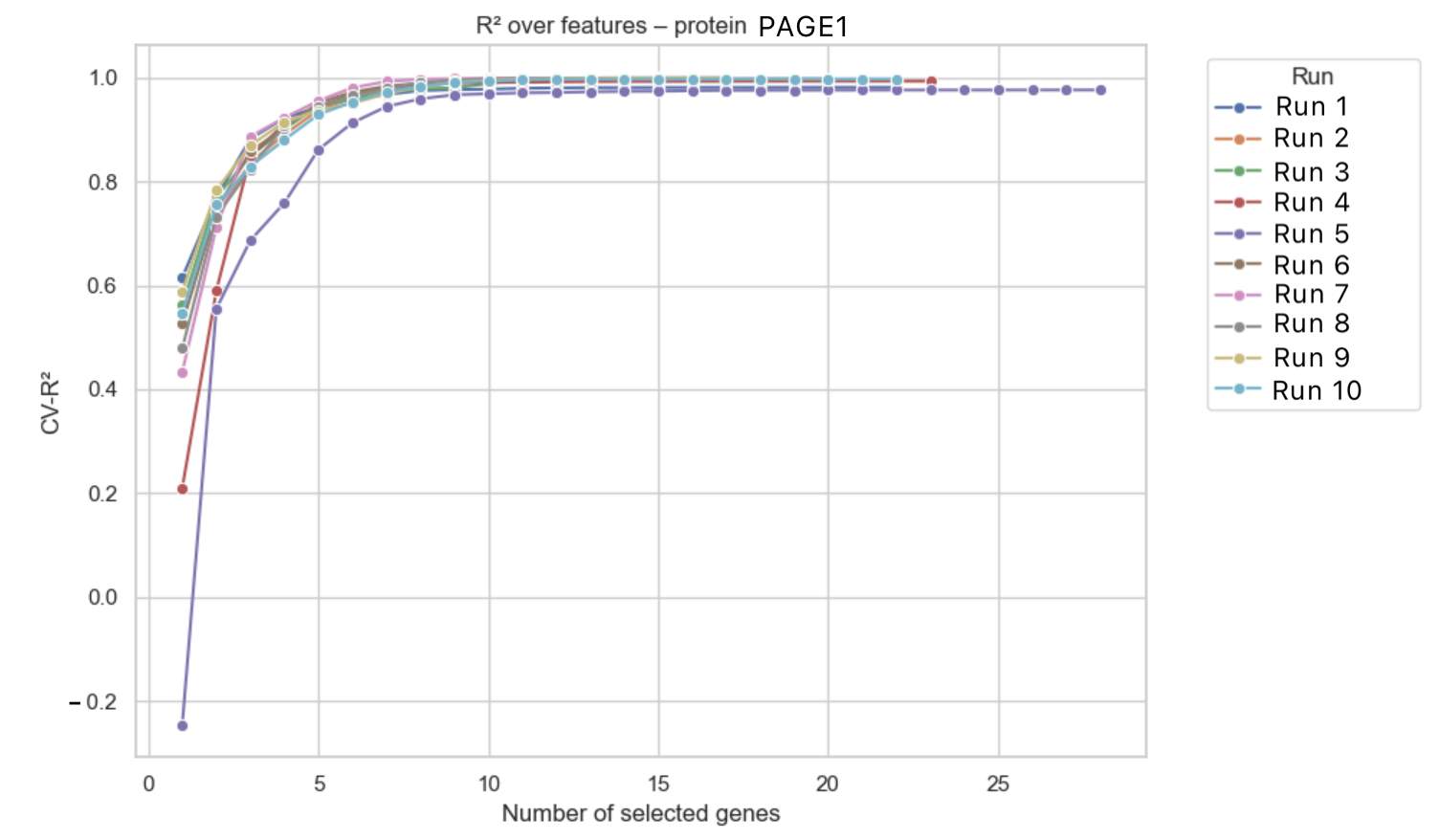}
\caption{$R^2$ improvement with number of selected features across all 10 runs for well-predictable protein \textbf{PAGE1}. The $R^2$ from 5-fold cross-validation improves only marginally, by selecting more than 10 features. Without selecting any features in addition to the cis-features, the model achieves $R^2$ scores between $-0.2$ and $0.6$. This indicates higher relevance of the cis-feature. Thus, the cis-feature of \textbf{PAGE1} seems to have quite high explanatory capacity as the run with the lowest $R^2$ can be perceived as an outlier, probably due to an unfortunate cross-validation splitting.}
    \label{fig:page1}
    
    \end{figure}

\begin{figure}[ht]

    \centering
    \includegraphics[width=\linewidth]{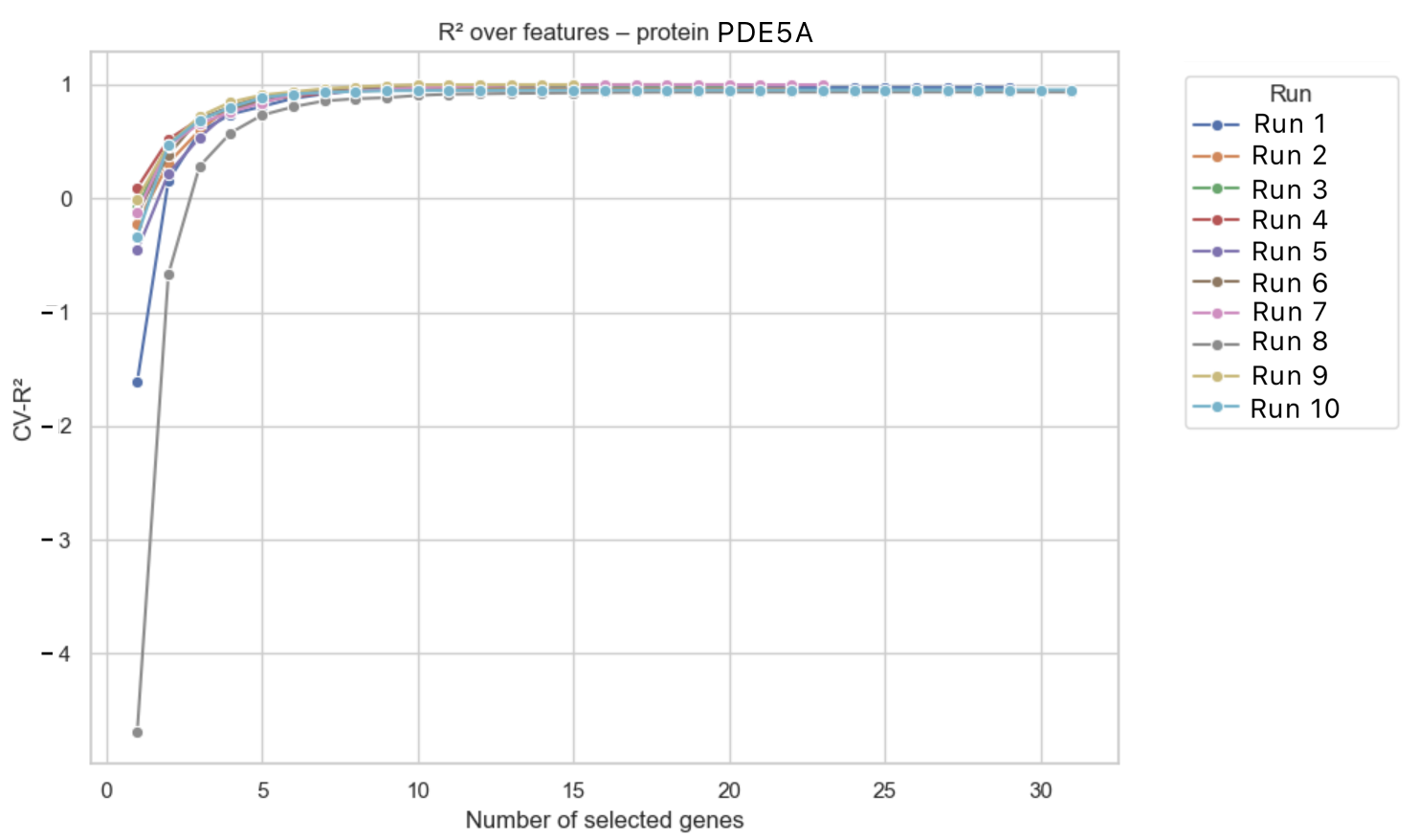}
\caption{$R^2$ improvement with number of selected features across all 10 runs for well-predictable protein \textbf{PDE5A}. Without selecting any features in addition to the cis-features, the model achieves $R^2$ scores below $0$ mostly. This indicates low explanatory relevance of the cis-feature, the one feature which needs to be involved by biological reasoning. Still, after approximately 15 selected features, the $R^2$ from 5-fold cross-validation reaches its optimum and stops improving significantly. This optimum is very high ($\geq 0.9$) and demonstrates the valuable contribution of our approach in explaining protein abundances through gene expression.}

    \label{fig:pde5a}
\end{figure}

For 8,166 proteins, we obtained interpretable sets of gene features with low to high selection stability. These gene–protein maps serve as potential signatures of transcriptional regulation. 
Here, we present the top-3 explainable proteins \textbf{PDE5A}, \textbf{PAGE1} and \textbf{XKR7} their top predictors and their selection frequencies.
(Table~\ref{tab:stable_features_proteins}).
Notably, we found no feature with \textit{stability score} $= 1.0 $ for any of the top-3 proteins except their \textit{cis-features}. Interestingly, we found only  one feature for each protein with a \textit{stability score} $\geq 0.6$.
Figures~\ref{fig:pde5a}, \ref{fig:page1} and \ref{fig:xkr7} show 10 feature selection runs with different random seeds for each of the top-3 performing proteins. In each of the plots, it is apparent that with the \textit{cis-feature} only the $R^2$ is far from optimal. Especially for \textbf{PDE5A} and \textbf{XKR7} the $R^2$, while using only the \textit{cis-feature}, barely outperforms the mean. For \textbf{PAGE1} the \textit{cis-feature} seems to be a better predictor for most random seeds. Still, by selecting only few features the $R^2$ quickly improves to be $\geq 0.8$. After adding between five and ten features, the corresponding curves seem to flatten.
For these three proteins we do not find any common stable predictors (Table~\ref{tab:stable_features_proteins}).

\subsubsection*{XKR7}

For the protein \textbf{XKR7}, we performed gene set enrichment analysis on Ridge-selected mRNA predictors (\textit{stability score} $\geq 0.2$, including \textbf{XKR7} itself). The \texttt{g:Profiler} analysis revealed significant enrichment in the \textit{canonical inflammasome complex} (GO:CC, $p = 0.004$), with two predictor genes mapping to this cellular component. Additionally, \textit{transcription factor binding motifs for HOXB5 and HOXB8} were significantly overrepresented ($p = 0.0035$ and $p = 0.0319$, respectively), present in 8 of the 13 genes analyzed. Notably, \textbf{XKR7 itself also contains these motifs}, suggesting that the protein may be part of a co-regulated transcriptional module linked to inflammatory or developmental signaling. These findings suggest that \textbf{XKR7} expression may be influenced by coordinated immunological activation and HOX transcriptional programs.

\subsubsection*{PAGE1}

For \textbf{PAGE1}, a cancer/testis antigen and the focus protein in this model, Ridge regression identified a set of 12 mRNA predictors with \textit{stability score} $\geq 0.2$. Enrichment analysis using \texttt{g:Profiler} yielded a single significant result: \textit{transcription factor binding motifs for ZNF549} ($p = 0.0166$), which were present in 8 out of 12 predictor genes. ZNF549 is a zinc finger transcription factor of currently unknown specificity, but its enriched binding pattern suggests that \textbf{PAGE1} and its top predictors may be under \textit{shared transcriptional regulation}, possibly in epigenetically active or developmentally programmed contexts. This result aligns with PAGE1’s known expression profile in germline and cancer cells.

\subsubsection*{PDE5A}

For the model predicting \textbf{PDE5A} protein expression, Ridge-selected mRNA features (\textit{stability score} $\geq 0.2$) were subjected to functional enrichment analysis. In contrast to \textbf{XKR7} and PAGE1, \texttt{g:Profiler} did not identify any significantly enriched functional categories or transcription factor motifs (FDR-adjusted $p \geq 0.05$). This suggests that the set of mRNA predictors for \textbf{PDE5A} is \textit{functionally heterogeneous}, lacking a dominant shared biological annotation. The absence of enrichment may reflect an indirect regulatory relationship, in which these transcripts serve as \textit{contextual or cell-type-specific markers} of \textbf{PDE5A} protein levels rather than being part of a coherent signaling pathway. However, the gene THEMIS2 emerged as a highly stable predictor (\textit{stability score} = 0.9), despite the absence of a known direct regulatory or functional link between these two genes in the current literature. This finding may reflect an indirect association—such as co-expression in specific cellular contexts, shared regulatory mechanisms, or context-specific pathway activity. Although the biological basis remains unclear, the high and consistent selection of THEMIS2 across multiple model runs suggests it may serve as a context-sensitive transcriptomic marker of \textbf{PDE5A} protein abundance.

\begin{table}[ht]
\centering
\caption{Stable features for the top-3 performing proteins \textbf{PDE5A}, \textbf{PAGE1} and \textbf{XKR7} and their \textit{stability scores} ($s \geq 0.2$)}
\label{tab:stable_features_proteins}
\begin{tabular}{lrlrlr}
\toprule
 \textbf{PDE5A} & &\textbf{PAGE1}&&\textbf{XKR7}& \\
\midrule 
Features&Score&Features&Score&Features&Score\\
\midrule 

 PDE5A-cis-f. & 1.0 & PAGE1-cis-f. & 1.0& XKR7-cis-f. & 1.0 \\
      THEMIS2     & 0.9 &CLCC1 & 0.7 &   CPNE2     & 0.9 \\
      KRT84     & 0.4 &ZFHX3 & 0.3  & DDX3X     & 0.4 \\
      NEFL     & 0.3 &LDHAL6B & 0.2& MANBAL     & 0.3 \\
      GIMAP6     & 0.3& LINC01600 & 0.2& OTC    & 0.2 \\
      SIRT3     & 0.3 &CYP19A1 & 0.2& GSTA3     & 0.2 \\
      RAB44     & 0.2 &MFF & 0.2     & CASP1     & 0.2 \\
      TFAP2B     & 0.2 &DAB2IP & 0.2 & GPR82     & 0.2 \\
      PLD4     & 0.2 &GUCA2B & 0.2  & CD48     & 0.2 \\
     HSP90AB2P     & 0.2 &GDF3 & 0.2 & NEK5      & 0.2 \\
      CFAP47     & 0.2 &SPINK6 & 0.2& MOB4     & 0.2 \\
      HBZ     & 0.2 &ACSM4 & 0.2& LYPLA2     & 0.2 \\
      HSP90AA2P      & 0.2 &NME2P1 & 0.2 & TMEM126A     & 0.2 \\
      SEPTIN4     & 0.2 &APOC2 & 0.2 & \\
      MUC12     & 0.2 &&&\\
      SPIB     & 0.2 &&&\\
      NYX     & 0.2 &&&\\
      CD34     & 0.2 &&&\\
     CASQ1     & 0.2 &&&\\
     MROH2B     & 0.2 &&&\\
      MTTP     & 0.2 &&&\\
      ANKLE2     & 0.2 &&&\\
      STYXL2     & 0.2 &&&\\
      PSMF1    & 0.2 &&&\\
      MYBPC2     & 0.2 &&&\\

\bottomrule
\end{tabular}
\end{table}

These results demonstrate that many proteins exhibit consistent transcriptomic predictability and that selected gene features reflect known biological groupings. These findings support the use of stable feature selection for mapping interpretable gene–protein relationships in large-scale omics datasets. The cis-features varying explanatory capacity across proteins is underlined by these findings.

\subsection{Frequently Selected Features}

To confirm the biological plausibility of our approach and to possibly discover genes that are consistently associated with protein abundance across cancer cell lines, we applied a repeated greedy feature selection with cross-validation and computed \textit{stability scores} for each gene across ten random seeds. Features with high stability are likely to represent biologically meaningful and generalizable gene–protein relationships.

\subsubsection{\textit{Stability Score} = 1.00}
\begin{figure}[ht!]
    \centering
    \includegraphics[width=0.75\linewidth]{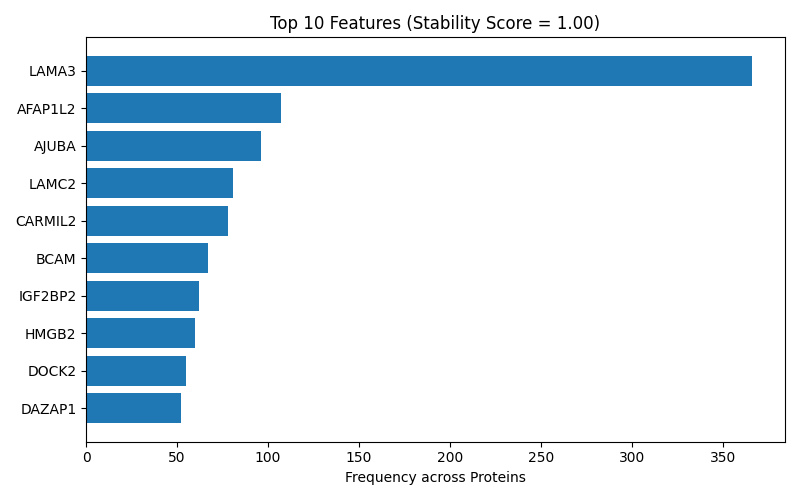}
    \caption{The most frequent genes across all proteins for \textit{stability score} $= 1.0$. Notably, \textbf{LAMA3} occurs 350 times and, thus, occurs more than three times as often as the next most frequent feature \textbf{AFAP1L2}. This indicates extraordinary robustness and relevance of \textbf{LAMA3}.}
    \label{fig:top_feats_100}

\end{figure}

At the highest stringency, we identified genes that were selected in \emph{all} runs for numerous proteins. Notably, \textbf{LAMA3} emerged as the most robust feature, appearing with perfect stability in over 350 protein models. This is by far the most frequently selected gene. Other consistently stable features included \textbf{AFAP1L2}, \textbf{AJUBA}, and \textbf{LAMC2} (Figure~\ref{fig:top_feats_100}). These genes may reflect broadly recurrent predictors of protein abundance across diverse cellular contexts.

\subsubsection{\textit{Stability Score} = 0.60}
\begin{figure}[ht!]
    \centering
    \includegraphics[width=0.75\linewidth]{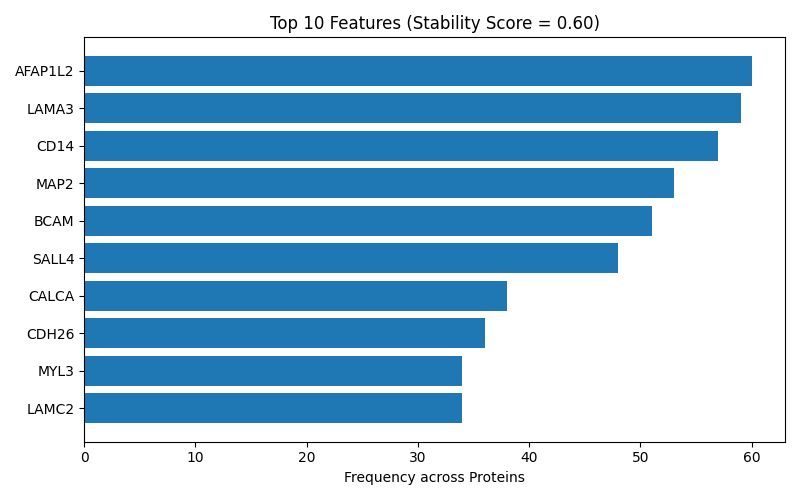}
    \caption{The most frequent genes across all proteins for \textit{stability score} $= 0.6$. Here, \textbf{AFAP1L2} and \textbf{LAMA3} lead the ranking while \textbf{MAP2} and \textbf{MYL3} are among the 10 most frequent features.}
    \label{fig:top_feats_060}
    \end{figure}

By relaxing the threshold to include moderately stable features, we observed partially different relevant predictors as top-10 predictors. For example, \textbf{LAMA3} is not the most frequent gene in contrast to higher \textit{stability scores}. \textbf{AFAP1L2}, \textbf{LAMA3}, and \textbf{CD14} remained among the most frequently selected, while genes such as \textbf{MAP2}, \textbf{BCAM}, and \textbf{SALL4} appeared more prominently (Figure~\ref{fig:top_feats_060}). These moderately stable features may capture regulatory signals that are shared across subsets of proteins or tissue types.

\subsubsection{\textit{Stability Score} = 0.20}
\begin{figure}[ht!]

    \centering
    \includegraphics[width=0.75\linewidth]{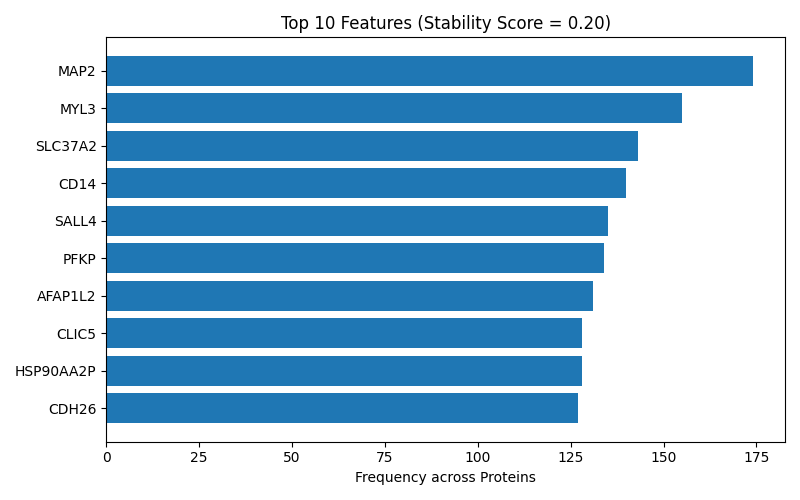}
    \caption{The most frequent genes across all proteins for \textit{stability score} $= 0.2$: Notably, all frequently selected genes occur similarly often. \textbf{MAP2} and \textbf{MYL3} are most frequent.}
    \label{fig:top_feats_020}
\end{figure}

At the lowest threshold, context-specific predictors became dominant. \textbf{MAP2}, \textbf{MYL3}, and \textbf{SLC37A2} were among the most frequently selected genes, albeit with greater variability across proteins (Figure~\ref{fig:top_feats_020}). Such genes may reflect lineage-specific or conditional expression–protein dependencies. \textbf{LAMA3} was not among the top-10 most frequent features anymore.

Across all thresholds, \textbf{AFAP1L2} consistently ranked among top features, suggesting broad relevance across the proteome. Still, for higher \textit{stability scores} \textbf{LAMA3} is the most frequently selected feature, thus, showing high relevance. Together, these results provide a structured overview of transcriptomic features most predictive of protein expression across cancer cell lines.


\section{Discussion}

Our study systematically investigates the stability and predictive utility of transcriptomic features in modeling protein abundance across cancer cell lines. 
By employing repeated greedy feature selection combined with cross-validated ridge regression, we provide a framework that yields interpretable and reproducible gene–protein associations on a proteome-wide scale.
We investigate the predictability of protein abundances through gene expression and identify relevant predictors per protein for more than 8,000 proteins.


We performed 10 independent runs of greedy feature selection using different random initializations and recorded the features selected per run. A stability score was assigned to each feature by counting the number of occurrences per feature selection run for the corresponding protein.
This quantitative \textit{stability score} introduces a key contribution of this work, capturing how consistently a transcriptomic feature is selected across multiple training splits and, thus, its robustness and relevance for a protein. This allows us to select robust features for protein abundance prediction with high reproducibility and interpretability. By this, we demonstrate the biological value of our approach combined with clean methodology.
Our results demonstrate that feature stability varies considerably across proteins and thresholds. At low thresholds (e.g., 0.1), a large number of features are retained, but many lack robustness. In contrast, high \textit{stability scores} (e.g., 0.6 or 1.0) drastically reduce the number of retained features but enrich for highly reproducible predictors.

This highlights a fundamental trade-off between comprehensiveness and robustness. The stability metric enables us to explicitly navigate this trade-off, selecting thresholds based on application-specific needs—for example, prioritizing reproducibility in biomarker discovery or allowing more flexibility in exploratory settings.
By incorporating stability selection across multiple random seeds, we identified gene features that are consistently selected across runs. This adds a layer of confidence to the interpretability of our results and addresses concerns about feature selection instability in high-dimensional data~\cite{10.1007/978-981-19-0151-5_26, Meinshausen2010, Haury2012}

These stable associations can guide downstream investigations into gene regulatory mechanisms or candidate biomarkers. Moreover, our results suggest that gene expression alone—despite its limitations—encodes biologically meaningful information about protein abundance for a substantial subset of proteins.


We found that for a substantial subset of proteins (between 222 and 3,555 depending on the \textit{stability score}), transcriptomic features alone enable accurate prediction of protein abundance ($R^2 > 0.5$), even under conservative stability constraints. Notably, \textit{stability score} 0.2 provided the best explainability for many proteins, although it also introduced noisy predictors for others. As stability increased, the number of low-performing models decreased, suggesting that higher stability also filters out spurious associations.

These findings show that gene expression data can be sufficient to predict protein abundances for a significant amount of proteins. However, the transcriptome–proteome coupling is rather moderate than very strong. While prior studies reported good performance for partial protein imputation within the same cell line~\cite{Barzine2020-fo} or on very few proteins only, we investigated more than 8,000 proteins across different cancer cell lines. For these proteins, our framework yields explicit, interpretable predictor sets per protein, which can serve as hypotheses for gene regulation or post-transcriptional buffering. This conclusion is further supported by additional benchmarking against the full transcriptome, which confirmed that our focused feature sets yield superior performance in the majority of cases (see Section~\ref{subsecFullTran}).


For many proteins, including top performers such as PDE5A, PAGE1, and \textbf{XKR7}, we identified small sets of robustly selected transcriptomic features that achieve high predictive performance. Interestingly, for PDE5A, the gene \textit{THEMIS2} emerged as a frequently selected and stable predictor, despite the absence of a known mechanistic link. This may suggest context-dependent co-expression, shared upstream regulation, or indirect biological coupling, a hypothesis that could be explored in follow-up studies.

Gene set enrichment analyses for \textbf{PAGE1} and \textbf{XKR7} revealed overrepresented transcription factor binding motifs (e.g., ZNF549, HOXB5/8), suggesting that groups of co-selected predictors may reflect regulatory modules rather than random associations. For PDE5A, no enrichment was detected, possibly indicating a more heterogeneous or indirect predictor set.

The most frequently and stably selected predictor genes in our models cluster into coherent biological modules likely involved in proteome regulation. \textbf{LAMA3} and \textbf{LAMC2}, both laminin-332 subunits, regulate basement membrane structure and integrin signaling, suggesting that Extra Cellular Matrix (ECM) remodeling broadly modulates protein abundance via PI3K/Akt/mTOR~\cite{Nonnast2025_LamininsCancer, Wu2025_LAMA3_ESCC}. \textbf{BCAM}, another laminin-binding receptor, supports this link~\cite{Sivakumar2025_BCAM}.

\textbf{AFAP1L2} and \textbf{AJUBA} are adaptor proteins implicated in cytoskeletal organization and mechanotransduction through PI3K/Akt and Hippo pathways, highlighting the role of actin signaling in protein-level control~\cite{Poosekeaw2021_XB130_CCA, Jagannathan2016_AJUBA_Hippo, McCormack2017_AJUBA_CdGAP}. \textbf{MAP2}, typically involved in neural and muscular cytoskeletal functions, may reflect lineage-specific or structural regulation~\cite{Wattanathamsan2022_MAPsMetastasis}.

\textbf{SALL4}, a stem cell factor with broad transcriptional impact, points to a developmental control layer~\cite{Zhang2018_SALL4}. Finally, \textbf{SLC37A2}, linking glycolytic flux and energy balance, suggests a metabolic component in protein regulation~\cite{Ji2025_SLC37A2}.

Together, these genes reflect mechanistic processes - ECM interaction, cytoskeletal signaling, development, and metabolism - that are consistently involved in transcript–protein associations in cancer cell lines.


While previous studies such as Barzine et al.~\cite{Barzine2020-fo} have explored transcriptome–proteome relationships in cancer cell lines, they did not systematically address the stability or interpretability of transcriptomic predictors. In contrast, our approach provides gene-level predictor sets per protein with quantified selection stability, enabling insights into potential regulatory signatures. This distinction allows not only performance assessment but also biological interpretation, which is essential for downstream functional studies or therapeutic target identification.


Our study focuses on ridge regression and greedy feature selection due to their interpretability and computational feasibility. While more complex models (e.g., Random Forests or neural networks) may improve raw predictive performance, they often lack transparency or require extensive training time. In future work, hybrid approaches that combine interpretability with non-linearity could be explored. This choice was further supported by a benchmark against Random Forest models, where ridge regression achieved better predictive performance in nearly all tested cases (see Section~\ref{subsecRandomForest}).

To assess the robustness of our findings beyond the training data, we performed an external validation using the independently generated CCLE proteomics dataset. Despite known variability and reproducibility challenges in proteomics technologies, a substantial number of GDSC-derived predictor sets remained predictive in CCLE. At a \textit{stability score} of 0.5, 3,679 of 7,155 protein models (51.4\%) achieved $PCC \geq 0.5$. These results support the reproducibility of the identified predictor sets across independent datasets and suggest that the observed associations are not solely dataset-specific.

Moreover, while our method captures statistical associations, it does not resolve causality. Experimental validation will be essential to confirm whether selected predictors reflect direct regulatory interactions or merely co-expression patterns. Finally, although our focus is on cancer cell lines, extending this analysis to primary tissues or other organisms could reveal conserved and context-specific transcriptome–proteome dynamics.

Additionally, our method currently treats proteins independently and does not exploit potential co-regulation or protein–protein interactions. Extending the framework to incorporate prior biological networks or integrate additional omics layers could further enhance interpretability and prediction performance.

For predictive tasks, we recommend using features obtained with a \textit{stability score} of 0.2, as these features yield the highest predictive accuracy. Features with a \textit{stability score} of 1.0, on the other hand, can be considered highly robust. These selections minimize noise and are therefore most suitable for biological reasoning and exploratory downstream analyses.
This robustness, however, must be interpreted in light of the redundancy inherent to biological systems. Biologically related genes often exhibit co-expression or shared functions, making them partially interchangeable in feature selection.
Given the sample size per model ($n = 940$), this interchangeability is expected and supports interpreting stability not as a measure of unique causal relevance, but as an indicator of robustness within groups of correlated predictors.

A practical compromise is achieved with a \textit{stability score} of 0.6, which combines predictive performance and robustness. At this threshold, stable features are identified for 7,564 proteins. Noisy predictors are largely avoided due to the consensus-based selection, where features are retained only if consistently chosen across the majority of runs. This trade-off makes \textit{stability score} 0.6 particularly well-suited for biologically grounded downstream applications of any kind.

\subsection{Conclusion}

Together, our results demonstrate that robust and interpretable gene–protein maps can be derived from transcriptomic data using systematic feature selection with stability analysis. These maps offer both predictive utility and biological insight, serving as a foundation for hypothesis generation in functional genomics and systems biology.

In future work, researchers benefit from our selected features for proteinwise investigations. Experimental validation may be able to show which of our stable features reflect causal relationships.
Ultimately, this study advances our understanding of translation by revealing stable, biologically meaningful links between the transcriptome and proteome.

\section{Methods}
\label{Methods}
\subsection{Data Description}

The data utilized in this study comprises gene expression and proteomics datasets. It is obtained from the Cell Model Passport (CMP)~\cite{cell_model_passport}. The CMP is a comprehensive database that provides molecular characterization of a wide range of cancer cell lines, facilitating integrative multi-omics research. The gene expression and protein intensity data from the CMP are publicly available. Detailed protocols for data generation, normalization, and quality control are provided in the CMP documentation, ensuring transparency and reproducibility of the data used in our study~\cite{cell_model_passport}.

The 37,602 gene expression values were sourced from RNA sequencing experiments. This data provides a quantitative measure of the transcriptional activity of genes across different cell lines. The dataset includes normalized expression levels for thousands of genes, measured in transcripts per million (TPM) to make cell lines comparable~\cite{cell_model_passport}:

\[TPM_i = \frac{q_i/l_i}{\sum_j(q_j/l_j)}*10^6\] 
\noindent
where \(q_i\) is the number of reads which are mapped to a transcript and \(l_i\) is the length of the transcript~\cite{Zhao2021}.

Proteomics data was obtained from DIA-MS. This data provides quantitative measurements of protein abundances across cell lines, representing the functional output of the genome. The proteomics dataset includes normalized protein intensity values for 8,449 proteins, processed through a series of steps including protein identification, quantification, and normalization to correct for systematic biases and technical variability. The resulting dataset offers a comprehensive view of protein expression profiles, facilitating the investigation of protein-level regulations~\cite{PROTEOMEDATAORIGINGONCALVES2022835}.

\subsection{Data Preprocessing}
Initial inspection revealed the presence of genes with entirely missing values throughout all samples, which were subsequently removed to ensure data integrity.
Still, after cleaning the dataset the protein expression included 38.65\% missing values while the feature set did not include any missing values anymore.
To integrate the gene expression and protein expression data, we aligned the datasets based on common cell line identifiers. Only cell lines with both gene expression and proteomics data available were included in the analysis. 
Then, we aligned the genes and proteins based on common names such that we obtained a final dataset of 940 cell lines with 8,423 gene expression values and 8,423  protein expression values per celllines (Figure~\ref{fig:translation}). This guarantees to have the \textit{cis-feature} for each protein.
Normalization of the feature set was carried out utilizing the \texttt{StandardScaler} from \texttt{scikit-learn}~\cite{SCIKIT_JMLR:v12:pedregosa11a}, which standardizes each value $x$ to the standard score $z$ for this value across all samples, thus, obtaining a standard score with a mean of zero and a standard deviation of one:

\[z = \frac{x-\mu}{\sigma}\] 
\noindent

where $\mu$ is the mean of the training samples and $\sigma$ is the standard deviation of the training samples. 

This step is critical for ensuring that all features contribute equally to the model's learning process.

\subsection{Greedy Feature Selection with Ridge Regression}

To identify transcriptomic features predictive of each protein, we implemented a greedy forward feature selection procedure. The method iteratively adds features from the full gene set that most improve the cross-validated $R^2$ score of a ridge regression model trained to predict the target protein.

The selection process was initialized with the transcript corresponding to the target protein itself (\textit{cis-feature}). In each step, the feature from the remaining pool that led to the largest increase in mean $R^2$ (evaluated via 5-fold cross-validation) was added to the model as indicated in Figure~\ref{fig:feature_selection}. 

To evaluate the predictive performance of our models, we used the coefficient of determination, denoted as $R^2$. It quantifies the proportion of variance in the dependent variable that is predictable from the independent variables. Mathematically, it is defined as:

\[R^2 = 1 - \frac{\sum_{i=1}^{n} (y_i - \hat{y}_i)^2}{\sum_{i=1}^{n} (y_i - \bar{y})^2}\] 
\noindent

Here, $y_i$ represents the observed protein expression values, $\hat{y}_i$ the predicted values from the model, $\bar{y}$ the mean of the observed values, and $n$ the number of samples.

The numerator represents the residual sum of squares (unexplained variance), while the denominator represents the total sum of squares (total variance). An $R^2$ value of 1 indicates perfect prediction, while a value of 0 indicates that the model does no better than simply predicting the mean. Negative values may occur when the model performs worse than the mean predictor.

As suggested by Guyon et al. the square root of the number of samples is a reliable orientation for the maximal number of features for machine learning models like Ridge~\cite{guyon2003introduction}. This strikes a balance between model complexity and statistical stability, especially when sample sizes are limited. By keeping the number of predictors low relative to the sample size, overfitting is reduced and generalization is improved.
As the $\sqrt{m} = 30.66$ where $m=940$ is the number of used samples in our data set and the \textit{cis-feature} is always part of the selected feature set, we defined the following two stopping criteria:
The selection stopped if either i) a maximum of $n-1 = 30$ features and the \textit{cis-feature} ($n = 31$ in total) was reached or if ii) the incremental gain in $R^2$ dropped below a threshold of 0.0001. Although this threshold lies below the typical variation expected from repeated cross-validation (standard error $\sim$0.001--0.002), it was deliberately chosen to allow the inclusion of weak but potentially meaningful predictors. To ensure robustness against overfitting due to random fluctuations, feature selection was repeated across 10 different random seeds. All selected features with their selection frequency (\textit{stability score}) were retained. This two-stage procedure allows for sensitive detection of relevant signals while maintaining model sparsity and reproducibility. Ridge regression with $\alpha = 2.0$ was employed in all runs, using implementations provided by scikit-learn.

\subsection*{Choice of Regression Model}

We employed Ridge regression as the core model for feature selection and predictive modeling. This choice was motivated by several characteristics of our dataset: a high number of features (genes), substantial multicollinearity, and relatively limited sample size (cancer cell lines). Ridge regression introduces an $L_2$ regularization penalty that stabilizes coefficient estimates while retaining the contribution of all features. This is particularly useful in biological systems where genes often function in co-regulated modules or pathways.

Compared to Lasso, which enforces sparsity through $L_1$ regularization and may arbitrarily select one gene among many correlated ones, Ridge regression distributes weights across correlated predictors. ElasticNet combines both penalties, but depending on the $\alpha$ parameter, it can still behave more similarly to Lasso and exclude informative features. For our goal — identifying stable, reproducible gene contributions to protein expression — Ridge provided the best trade-off between interpretability and stability, especially in combination with cross-validation and multiple random seeds.

This decision is in line with previous work in bioinformatics, where Ridge regression has been shown to outperform sparse models in high-dimensional, collinear settings with limited samples~\cite{sokolov2016}.

\subsection{Stability Score}
\label{sec:StabilityScore}
To assess the robustness of selected features, we repeated the feature selection procedure across multiple random seeds. For each protein, we performed 10 independent runs of greedy feature selection using different random initializations (random seeds) and recorded the features selected in each run.

For every gene feature, we then computed a \textit{stability score} $S_f$ defined as the proportion of runs in which the feature was selected:

\[
S_f = \frac{T_f}{r}\]
where $T_f$ is the number of times feature $f$ was selected across $r$ repetitions.

All selected features and \textit{stability score}s were retained for further analysis.

\subsection{Evaluation of Explainability}

To evaluate how well transcriptomic features explain individual proteins and which \textit{stability score} is the most reliable for predictions, we trained ridge regression models for each \textit{stability score} per protein and assessed performance via 5-fold cross-validation with new random seed and the already selected features. We report the cross-validated $R^2$ scores for each protein as a measure of in-sample predictability.

To evaluate the generalizability of the identified gene–protein associations, we conducted an external validation using the CCLE proteomics dataset~\cite{CCLEBarretina2012}. For each \textit{stability score}, we transferred the predictor sets (selected features) derived from our GDSC2-based framework and retrained ridge regression models using CCLE transcriptomic and proteomic data. Model performance was assessed by computing Pearson correlation coefficients (PCC) between predicted and observed protein levels. This setup allows us to test whether the GDSC2-derived associations remain predictive in an independent dataset. This strategy ensures that model weights are optimized specifically for the CCLE dataset while maintaining the feature selection logic derived from GDSC2.

\subsection{Benchmarking Experiments}

To validate our modeling framework, we conducted two additional benchmarking experiments. First, we compared models trained on the selected 8,423 focus-genes with models using the full transcriptome of 37,602 genes. This assessed whether restricting the feature set compromises predictive performance (see Section~\ref{subsecFullTran}).

Second, we compared our ridge regression approach to Random Forest models as a representative non-linear baseline. Both comparisons were carried out on a randomly selected subset of 100 proteins using identical cross-validation procedures. The goal was to evaluate whether our method offers competitive performance while preserving interpretability (see Section~\ref{subsecRandomForest}).

\section*{Acknowledgements}

This work was funded by the European Union under the Horizon Europe
grant 101186829 (CancerScan).

\section*{Availability of source code and requirements}

\begin{itemize}
\item Project name: Retracing the Process of Translation
\item Project home page: \url{https://github.com/JHelge/ReTrans.git}
\item Operating system(s): Platform independent
\item Programming language: Python, Bash
\item License: MIT

\end{itemize}

\section*{Competing Interests}

The author(s) declare that they have no competing interests

\section*{Use of AI Tools and Technologies in Writing}

During the preparation of this work the authors used ChatGPT~\cite{ChatGPT} in order to assist in refining the clarity and correctness of the text, particularly with respect to grammar, vocabulary usage, and stylistic consistency. The tool was utilized as a supplementary aid to enhance the overall readability and coherence of the manuscript. After using this tool/service, the authors reviewed and edited the content as needed and take full responsibility for the content of the publication.

\bibliographystyle{unsrt}  
\bibliography{templateArxiv}

\end{document}